\documentclass[12pt,letterpaper]{article}
\usepackage{graphicx,lineno,amsmath,amssymb} 
\usepackage{fullpage,natbib}
\begin{document}
\title{Multiporosity Flow in \\Fractured Low-Permeability Rocks}
\author{Kristopher L. Kuhlman$^1$,\\Bwalya Malama$^2$,\\Jason E. Heath$^3$\\
\\
$^1$Applied Systems Analysis \& Research Dept., Sandia National Laboratories\\
$^2$Dept. of Natural Resources Management \& Environmental Sciences,\\California Polytechnic State University at San Luis Obispo\\
$^3$Geomechanics Dept., Sandia National Laboratories}

\maketitle
\begin{abstract}
 A multiporosity extension of classical double and triple porosity fractured rock flow models for slightly compressible fluids is presented. The multiporosity model is an adaptation of the multirate solute transport model of \cite{haggerty1995} to viscous flow in fractured rock reservoirs.  It is a generalization of both  pseudo-steady-state and transient interporosity flow double porosity models. The model includes a fracture continuum and an overlapping distribution of multiple rock matrix continua, whose fracture-matrix exchange coefficients are specified through a discrete probability mass function.  Semi-analytical cylindrically symmetric solutions to the multiporosity mathematical model are developed using the Laplace transform to illustrate its behavior.  The multiporosity model presented here is conceptually simple, yet flexible enough to simulate common conceptualizations of double and triple porosity flow. This combination of generality and simplicity makes the multiporosity model a good choice for flow in low-permeability fractured rocks. 
\end{abstract}
\section{Introduction}
The flow of slightly compressible fluids through fractured rocks is of fundamental importance to groundwater and hydrocarbon production and effective isolation of radioactive waste and CO$_2$. In low-permeability rocks, fractures are often the primary source of bulk permeability. Reservoir and storage rocks often include multiple overlapping and intersecting fracture sets. Discrete representations of fractures in numerical models is possible if information is available about fracture spacing, orientation, and aperture. Typically, spatial distributions of these data are not available, and fractured rocks must be approximated as porous media \citep{gringarten1982flow,sahimi11}.  Multiple porosity types can simultaneously participate in flow through fractured and unfractured rocks. Shales have macroscopic porosity associated with clastic particles and microscopic porosity associated with their organic fraction \citep{akkutlu12}. The Culebra Dolomite has illustrated effects of multiple porosity types (fractures, vugs, and intergranular porosity) during flow and solute transport at both the field and laboratory scale \citep{mckenna2001,malama13core}. The effect of multiple types and scales of porosities can be masked by or mistaken for the effects of multiple types and scales of heterogeneity \citep{altman02}.

Dual or multiple porosity models are common simplified conceptualiztaions of this complex reality of interacting porosity types and the spatially heterogeneous nature of rocks. The treatment of fractured rock as a system of interacting and overlapping continua has been used successfully in its different forms (e.g., \citealp{gringarten1982flow}; \citealp{aguilera80}; \citealp{streltsova88}; \citealp{chen1989transient}; \citealp{daprat90}; or \citealp{bourdet02}) since its first introduction by \cite{barenblatt60a}.  The conceptualization is based on a low-permeability but high storage-capacity rock matrix drained via natural and man-made high-permeability but low storage-capacity fractures.

Slightly compressible fluids include water, brine, oil, and other liquids for which density can be assumed constant, allowing a formulation in terms of head (rather than pressure). Gases can be treated as a slightly compressible fluids after a transformation to create a pseudo-pressure and pseudo-time, which accounts for pressure-dependence of fluid properties  \citep{friedmann58,al-hussainy66}. Slightly compressible flow models can also approximate more complex non-linear flow (i.e., multiphase flow, time-variable permeability, or gas desorption) by transforming observations into one of several possible integrated pseudo variables (see review by \citealp{clarkson13review}).  Flow prediction for slightly compressible fluids in fractured low-permeability rocks at a macroscopic (wellbore or reservoir) scale is of great importance to applications in groundwater supply, hydrocarbon production, and underground sequestration of nuclear waste or CO$_2$.  

Fractured reservoirs exhibit pressure drawdown due to production at a specified pressure or production at a specified flowrate (e.g., during recovery or shut-in), in a manner characteristic of multiple interacting porosities (e.g., \citealp{crawford76,gringarten1982flow,moench84}). Fractures provide high-permeability pathways, often orders of magnitude more permeable than the unfractured rock itself, but comprise only a small portion of the rock volume. Natural fracture porosity is often less than $0.1\%$, but the fracture network is a much more efficient fluid conductor per unit porosity than the matrix \citep{streltsova88}.  Several well-known double porosity conceptualizations for ``uniformly'' or ``naturally'' fractured reservoirs have been developed (e.g., \citealp{barenblatt60b}; \citealp{warren1963behavior}; and \citealp{kazemi1969spe}), which approximate the matrix as being storage only (no advection outside of fractures). These solutions represent the flow domain with overlapping fracture and matrix continua.  These models assume the domain of interest is large enough and the fracture density is uniform enough to treat both fractures and matrix as continua. Fractures are the main source of permeability and spatial connectivity, while the matrix is the main source of storage. Triple porosity solutions with a single fracture continuum and two matrix continua (only connected through the fracture porosity) are a logical extension of dual porosity (e.g., \citealp{clossman75}; \citealp{ci1981exact}; \citealp{al-ahmadi11}; and \citealp{tivayanonda12}). 

Beginning from the system proposed by \citet{barenblatt60a}, but equally allowing advection through fractures and matrix results in the dual permeability model, which requires a coupled set of governing equations and boundary conditions. These were first solved analytically by \cite{chen80} in terms of quasi-Bessel functions, but this solution has not seen wide use. Although the dual-permability approach is more physically realistic than the simplified dual-porosity approach, when matrix permeability is much smaller than the fractures, dual porosity is an adequate approximation \citep{chen1989transient}. The multiple interacting continua (MINC) approach taken by TOUGH2 is a numerical approach capable of simulating both dual-porosity and dual-permeability systems \citep{pruess85,pruess99}.   

Many analytical solutions for variations on double and triple porosity conceptual models have been developed \citep{moench84,chen1989transient,daprat90}, with different configurations and relationships between fracture and matrix continua, but a generalized solution is needed.  We present multiporosity (the name coming from multirate and dual-porosity) as a generalization and extension of the \citet{warren1963behavior} solution to any number of matrix continua interacting with a fracture continuum.  Multiporosity is an adaptation of the multirate solute transport theory to viscous flow, for computing pressure-driven flow through low-permeability rocks with heterogeneity and multiple porosities.  The multirate (i.e., multiple reaction rates) conceptual model for solute transport \citep{haggerty1995,haggerty1998} has been successfully used to simulate diffusion of solutes from fast-flowing fractures into a distribution of diffusion-dominated matrix block sizes \citep{haggerty2000,haggerty2001,mckenna2001,malama13core}. 

\section{Multiporosity Model}
It is useful to conceptualize slightly compressible flow in uniformly fractured domains as diffusion in a porous medium.  Some deviations from this ideal behavior can be accommodated through use of pseudo-time and -pressure \citep{clarkson13review}.
 
\subsection{Conceptualization}
Two primary flow conceptualizations used in slightly compressible dual-porosity systems are:

\begin{enumerate}
\item The pseudo-steady-state \cite{warren1963behavior} (WR) interporosity flow conceptualization (similar to \citealp{barenblatt60b}) assumes flow from matrix blocks is proportional to the pressure difference  between the fracture and the average matrix pressures. 
\item The \citet{kazemi1969spe} (KZ) transient interporosity flow conceptualization allows transient diffusion from the fracture to the matrix, and couples flow in the fracture and matrix domains through a source term proportional to flux in the fracture flow governing equation. 
\end{enumerate}

The WR matrix flow conceptualization is much simpler than the KZ conceptualization, but leads to analytical flow solutions more readily than the KZ approach (e.g., see review by \citealp{chen1989transient}). The KZ approach has been solved using finite differences \citep{kazemi1969spe}, finite volumes \citep{pruess99}, and various analytical (e.g., \citealp{raghavan1983new}, \citealp{chen1985pressure}, or \citealp{ozkan1987unsteady}) and approximate \citep{walton13} approaches.  The WR solution produces physically unrealistic pressure transient solutions during the transition from early fracture-dominated flow to later matrix-dominated flow \citep{gringarten1982flow,moench84}.  It produces a nearly flat transition between early fracture flow and late-time matrix flow on a semi-log plot, equivalent to a vanishing log time derivative -- $\partial / \partial (\ln t)$.  \citet{moench84} showed the WR model could be improved by adding a fracture skin, which delayed communication between the fracture and matrix. The WR approach assumes pseudo-steady-state flow between fracture and matrix, which may not occur physically until late time in low-permeability rocks.  The KZ model produces a half-slope transition between the early and late-time flow, which is more in agreement with typical field observations \citep{moench84}.

The multiporosity approach is a generalization of the WR double-porosity model and the pseudo-steady-state triple-porosity model of \cite{clossman75}.  Triple-porosity models can represent two systems of natural preexisting micro- and macro-fractures \citep{al-ahmadi11}.  The multiporosity model is presented here as a spatial distribution of natural fractures and matrix materials (i.e., matrix heterogeneity).

When moving between models with different numbers of porosities (e.g., double- or triple-porosity compared to single-porosity models), a model with more parameters is typically more flexible and able to fit a wider range of observed behaviors, but more free parameters must be estimated.  Data collected from typical low-permeability wells are inadequate to uniquely constrain models with many estimable parameters.  To constrain the number of free parameters in the multiporosity model, the interporosity flow parameters can be specified with a distribution function (e.g., \cite{ranjbar2012one}).  \cite{haggerty1995} derived a distribution comprised of an infinite series of porosities.  Their distribution is equivalent to a distribution of pseudo-steady-state WR matrix porosities that behave in total as a transient KZ-type system.  The multiporosity model can be made equivalent to diffusion into a slab (i.e., the KZ conceptualization), cylinder, or sphere.  During parameter estimation, the multiporosity model can be matched to data either with flexible but parsimonious property distributions (e.g., lognormal or beta), or with an arbitrary number of individual porosities and their associated parameters.   The multiporosity distribution can be flexible, but with certain distributions it can also  be shown to be a generalization of two well-known physically based end members. 

Porosity distributions have been utilized in some dual-porosity solutions based upon different block-size distribution function (e.g., \citealp{mcguinness86}; \citealp{chen1989transient}; and \citealp{ranjbar2012one}), which is similar to the approach taken here.  Through parameter estimation, matrix block distributions can provide a fracture distribution, which is much more flexible than typically double or triple porosity models that assume more uniformly spaced fracture distributions.  Existing block-size distribution solutions have focused primarily on the derivation of shape factors.  We show how certain distributions of porosities lead to the well-known special cases of WR and KZ double-porosity flow.

A multiporosity conceptual model for flow has two possible meanings. Multiporosity can represent a single matrix continuum with the inter-porosity exchange coefficient between the fracture and matrix continua treated as heterogeneous (i.e., a random variable). This is a simplification of small-scale spatial heterogeneity  (e.g., \citealp{haggerty1995}; \citealp{haggerty1998}; or \citealp{akkutlu12}).  Alternatively, it can represent a set of multiple physically-distinct matrix continua with the porosity and permeability of each as random variables.  Our discussion takes the former approach, we believe it to be the most physically realistic.

\subsection{Mathematical Model}
We present the multiporosity flow model, which is a logical extension of the WR conceptualization \citep{warren1963behavior}, to an arbitrary number of pseudo-steady-state matrix domains.  The model is an adaptation of the multirate advection-dispersion solute transport solution \citep{haggerty1995,haggerty1998}  to viscous flow. It conceptualizes random porosity spatial variability in rock matrix as a random distribution of uniform matrix porosities communicating with the primary fracture porosity.

The governing equation for pressure head drawdown $\Delta p(\mathbf{x},t) = p_0(\mathbf{x}) - p(\mathbf{x},t)$ ($p_0$ is initial pressure head) in the fracture continuum is
\begin{equation}
  \label{eq:multirate-frac-flow1}
  \phi_f c_f \frac{\partial \left(\Delta p_f \right)}{\partial t} + \sum_{j=0}^{N} \phi_j c_j \chi_j \frac{\partial \left(\Delta p_j \right) }{\partial t}  = \frac{k_f}{\mu} \nabla^2 \left( \Delta p_f \right), 
\end{equation}
where $t$ is time, index $j$ and subscript $f$ denote quantities related to the $j$th matrix and fracture continua, the sum is across $N$ matrix porosities ($N$ may be infinite), $\phi$ is dimensionless porosity, $c$ is a compressibility or storage coefficient, $k$ is permeability, $\mu$ is fluid viscosity, and $\chi$ is a dimensionless probability mass function (PMF -- the discrete form of a probability density function) of interporosity exchange coefficients.  See Table~\ref{tab:nomenclature} for a summary of physical quantities and their units.  

In this multiporosity conceptualization, matrix properties and dependent variables are implicitly discrete functions of the distribution of matrix continua index. The properties controlling matrix-fracture fluid exchange across a potentially infinite number of matrix-fracture interfaces can be considered a random variable, but the resulting governing equations are deterministic because only their sum or bulk behavior appears in (\ref{eq:multirate-frac-flow1}). 

Flow in each of the matrix domains is generally governed by the diffusion equation, viz.  
\begin{equation}
  \label{eq:multirate-matrix-gen}
  \phi_j c_j \frac{\partial \left( \Delta p_j \right)}{\partial t} = \frac{k_j}{\mu} \nabla^2 \left( \Delta p_j \right) \qquad j=1,\dots,N;
\end{equation}
the governing equations will be non-dimensionalized and solved using the Laplace transform.

The governing multiporosity fracture flow equation (\ref{eq:multirate-frac-flow1}) can be non-dimensionalized by dividing through by the total formation compressible storage $\phi c = \phi_f c_f + \sum_{j=1}^{N}\phi_j c_j$,  and a characteristic pressure $P_c$, to produce the dimensionless fracture flow equation
\begin{equation}
  \label{eq:multirate-frac-flow-nond}
  \omega_f  \frac{\partial \psi_f}{\partial t_D} + \sum_{j=1}^{N} \omega_j \chi_j \frac{\partial \psi_j }{\partial t_D} =  \nabla_D^2 \psi_f,
\end{equation}
where $\psi_{\ell} = \Delta p_{\ell}/P_c$ $\left(\ell \in \lbrace f,j \rbrace \right)$ is dimensionless pressure change, $\omega_{\ell}=\phi_\ell c_\ell/\left(\phi c \right)$ is the fractional storage of an individual continuum ($0 \le \omega_{\ell} \le 1$ and $\sum \omega = 1$), $\nabla^2_D/L_c^2 =\nabla^2 $ is the dimensionless Laplacian, $L_c$ is a characteristic length (often defined as the pumping well radius, $r_w$),  $t_D=t/T_c$, and $T_c=L_c^2 \mu \phi c/k_f$ is a characteristic time.  Alternatively, $\omega_\ell$ can be related to the volume-weighted storage coefficient commonly used in hydrogeology, $\omega_\ell = S_{s\ell} V_\ell/\left( S_{sf}V_f + \sum_j S_{sj}V_j\right)$, where $V$ is a fraction of the total volume \citep{gringarten1982flow,moench84}.  Table~\ref{tab:dimensionless} defines dimensionless quantities for the current problem.

The matrix flow equation (\ref{eq:multirate-matrix-gen}) can analogously be non-dimensionalized into
\begin{equation}
  \label{eq:multirate-matrix-gen-nond}
  \omega_j  \frac{\partial \psi_j}{\partial t_D} = \kappa_j \nabla_D^2 \psi_j \qquad j=1,\dots,N,
\end{equation}
where $\kappa_j=k_j/k_f$ is the $j$th matrix-to-fracture permeability ratio.  

\begin{figure}  % Figure 1
  \noindent\includegraphics[width=0.75\textwidth]{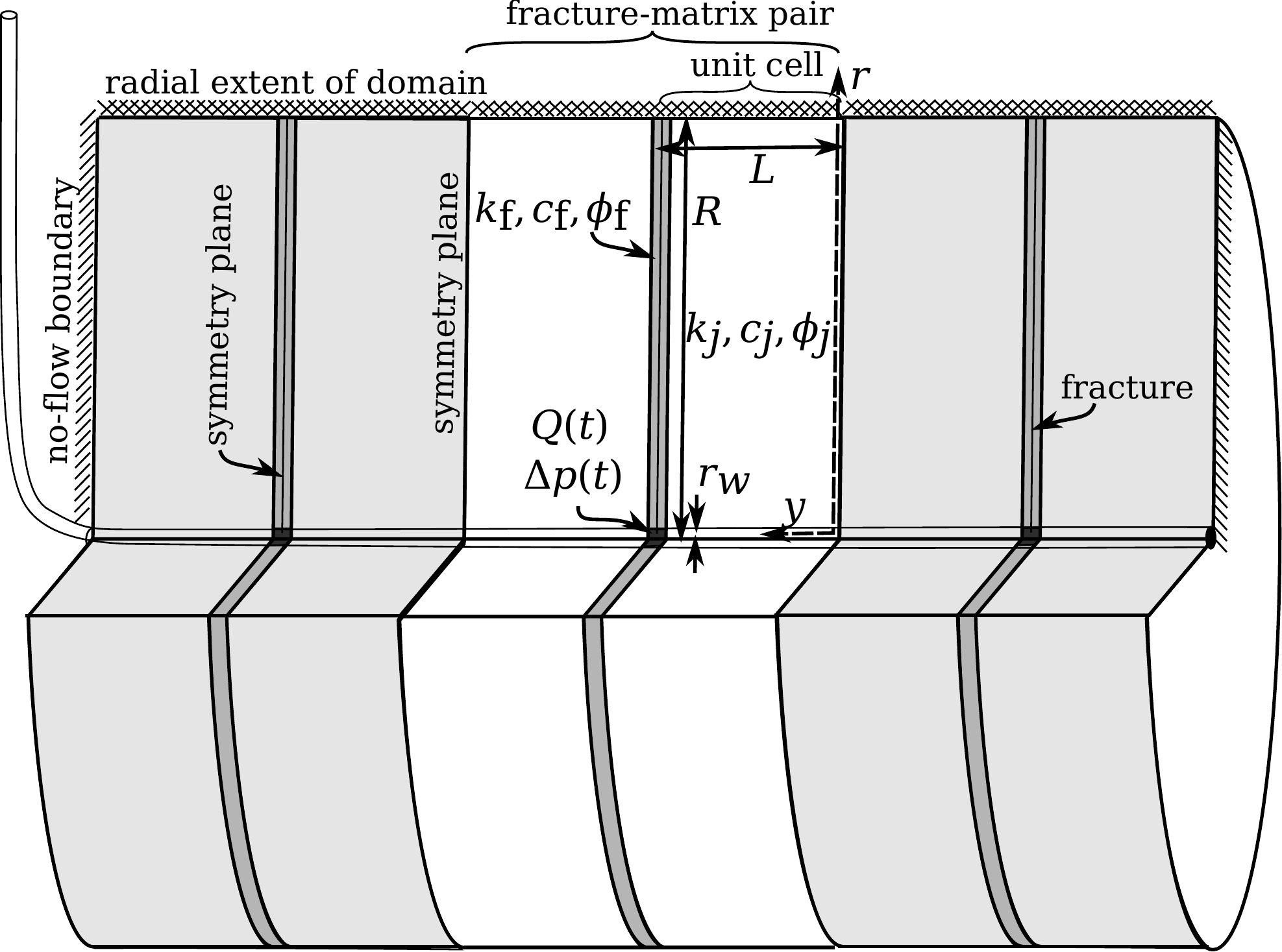}
  \caption{Equivalent layered representation of multiporosity problem assuming one-dimensional slab-type matrix geometry. Matrix is white cylindrical volume on either side of fractures. Additional fracture-matrix pairs illustrated in gray on either side of primary pair.}
  \label{fig:kazemi-slab}
\end{figure}

To simplify from the transient  to the pseudo-steady-state interporosity flow conceptualization we use average matrix pressure by integrating the matrix governing equation (\ref{eq:multirate-matrix-gen-nond}) across the matrix domain.   We assume matrix blocks are comprised of one-dimensional slabs (i.e., rectangular radially symmetric blocks with flow perpendicular to fractures, see Figure~\ref{fig:kazemi-slab}). The layered idealization in Figure~\ref{fig:kazemi-slab} results in an equivalent solution to the dual porosity conceptualization when continua exist at each physical location \citep{streltsova88}.  We choose slabs for simplicity; other possible matrix block geometries (e.g., spheres or cylinders) do not result in markedly different predicted results for the double porosity conceptualization (e.g., \citealp{gringarten1982flow,moghadam2010dual}). Integrating  (\ref{eq:multirate-matrix-gen-nond}) across a one-dimensional slab results in
\begin{equation}
  \label{eq:multirate-matrix-integrated}
  \frac{\partial \left \langle \psi_j \right \rangle}{\partial t_D} = \frac{\kappa_j}{L_{D} \omega_j } \left[ \left . \frac{\partial \psi_j}{\partial y_{D}} \right |_{y_{D}=L_{D}} - \left . \frac{\partial \psi_j}{\partial y_D} \right|_{y_D=0} \right],
\end{equation}
where $L_{D}=L/L_c$ is the dimensionless half-distance between evenly spaced fractures, $y_D=y/L_c$ is the dimensionless matrix space coordinate perpendicular to the fracture ($0 \le y_D \le L_{D}$), and 
\begin{equation*}
\left \langle \psi_j \right \rangle(t) = \frac{1}{L_{D}} \int_{0}^{L_{D}} \psi_j(x,t) \;\mathrm{d}x
\end{equation*} 
is the spatially averaged dimensionless change in matrix pressure.  The first term on the right-hand-side of (\ref{eq:multirate-matrix-integrated}) is flux at the matrix-fracture interface, while the second term is flux at the center of the block. The latter vanishes identically by symmetry.

The two no-flow boundaries perpendicular to the borehole in Figure~\ref{fig:kazemi-slab} are symmetry boundary conditions, which define the unit cell represented by the model given here.  In the case of a systematically fractured horizontal well one plane is in the midplane of the fracture and the other halfway between equally spaced fractures. The cylindrical boundary parallel to the borehole represents the edge of the reservoir -- potentially at infinite radial distance.  At this far edge of the domain we implement a general linear Type-III boundary condition.  

Continuing with the WR pseudo-steady-state interporosity flow assumption, flux from each matrix continuum to the fracture continuum is proportional to the difference between the dimensionless fracture and average matrix pressure changes,
\begin{equation}
  \label{eq:flux-expression}
  \left . \frac{\partial \psi_j}{\partial y_D} \right |_{y_D=L_{D}} = \frac{\epsilon_j}{L_{D}} \left[ \psi_f - \left \langle \psi_j \right \rangle \right] \qquad j=1,\dots,N,
\end{equation}
where $\epsilon_j$ is a dimensionless constant of proportionality. Substituting this expression for flux (\ref{eq:flux-expression}) into the integrated matrix flow equation (\ref{eq:multirate-matrix-integrated}) leads to
\begin{equation}
  \label{eq:matrix-WR-form}
  \frac{\partial \left \langle \psi_j \right \rangle}{\partial t_D} = \frac{\epsilon_j \kappa_j}{L_{D}^2 \omega_j} \left[ \psi_f - \left \langle \psi_j \right \rangle \right]  \qquad j=1,\dots,N.
\end{equation}
Comparing the $j=1$ matrix porosity from (\ref{eq:matrix-WR-form})  to the standard WR solution for matrix flow in a double-porosity system \cite[Eqn.~11]{warren1963behavior}, results in the equivalence
\begin{equation}
  \label{eq:WR-comparison}
  \frac{\alpha r_w^2 k_m}{(1-\omega) k_f} = \frac{\epsilon_1 k_1 L_c^2}{L^2 \omega_1 k_f},
\end{equation}
which suggests $\epsilon_1 = L_{D}^2 \alpha$, where $\alpha$ is WR's shape parameter [L$^{-2}$], $k_m=k_1$ is the WR matrix permeability, $\omega=\omega_f$, $\omega_1 = 1-\omega$, and WR used $L_c = r_w$ as a characteristic length. 

We choose the group
\begin{equation*}
  u_j=\frac{\epsilon_j \kappa_j }{\omega_j}
\end{equation*}
to characterize the flow problem across the distribution of matrix continua ($0 \le u_j < \infty$). Taking the Laplace transform of (\ref{eq:matrix-WR-form}) results in
\begin{equation}
  \label{eq:multirate-matrix-L}
  \left \langle \bar{\psi}_j \right \rangle s  = u_j \left[ \bar{\psi}_f - \left \langle \bar{\psi}_j \right \rangle \right]  \qquad j=1,\dots,N
\end{equation}
where $s$ is the dimensionless Laplace transform parameter and an overbar indicates a transformed dependent variable, i.e.,  $\bar{f}=\int_0^{\infty} e^{-st_D} f(t_D)\; \mathrm{d}t_D$. The averaged change in dimensionless matrix pressure due to changes in the fracture pressure is 
\begin{equation*}
  \left \langle \bar{\psi}_j \right \rangle = \frac{u_j \bar{\psi}_f}{s + u_j }.
\end{equation*} 
This can be substituted into the Laplace-transformed form of (\ref{eq:multirate-frac-flow-nond}) after similarly integrating \eqref{eq:multirate-frac-flow-nond} across the matrix blocks (equivalent to replacing $\psi_j$ with $\left \langle \bar{\psi}_j \right \rangle$), resulting in 
\begin{equation}
  \label{eq:multirate-nonD-fracture}
     s \omega_f \bar{\psi}_f  + s \omega_f \sum_{j=1}^{N} \left[ \beta_j \chi_j \frac{u_j }{s + u_j} \right] \bar{\psi}_f =  \nabla_D^2 \bar{\psi}_f,
\end{equation}
where $\beta_j=\omega_j/\omega_f$ is a dimensionless matrix/fracture storage capacity ratio.
 
Equation (\ref{eq:multirate-nonD-fracture}) can be further simplified into
\begin{equation}
  \label{eq:multirate-nonD-fracture2}
    \nabla_D^2 \bar{\psi}_f - \bar{\psi}_f \omega_f s \left( 1 + \bar{g} \right) = 0,
\end{equation}
where
\begin{equation}
\bar{g}=\sum_{j=1}^{N}  \frac{ \hat{\chi}_j u_j}{s + u_j},
\label{eq:g}
\end{equation}
is the matrix memory kernel \citep{haggerty1995} and $\hat{\chi}_j=\chi_j\beta_j$ is a scaled PMF.

\subsection{Distributions}
To compute a solution to (\ref{eq:multirate-nonD-fracture2}), the matrix memory kernel must be specified.   Any valid PMF can be specified for $\chi_j$, but we present three special cases.  The multiporsity solution simplifies to the dual-porosity WR solution through
\begin{equation}
  \label{eq:WR-g}
  \bar{g}_\mathrm{WR} = \frac{\left(\frac{1}{\omega} - 1\right) \frac{\lambda}{1-\omega}}{s + \frac{\lambda}{1-\omega}}, 
\end{equation}
which is $u_1=\lambda/(1-\omega)=\lambda/\omega_1 = \alpha r_w^2 \kappa/\omega_1$, $\chi_1=1$, and $N=1$.  Here $\lambda=\alpha\kappa r_w^2$ is WR's dimensionless interporosity flow parameter.

Similarly, the multiporosity solution is equivalent to the pseudo-steady-state triple-porosity solution of \cite{clossman75} or \cite{ci1981exact} using (\ref{eq:g}),  
\begin{align*}
u_j & = \kappa_j/\omega_j\\
\chi_j& =\frac{\gamma_j(1-\gamma)}{\gamma}
\end{align*} 
and $N=2$, where $\gamma$ is the bulk volume fraction of fissures, and $\gamma_{\{1,2\}}$ are the volume fraction of good and poor rock in the matrix \citep{odeh65,clossman75}. 

\citet[Table 1]{haggerty1995} presented infinite discrete distributions which make the overall  multiporosity system (comprised of a sum of WR pseudo-steady-state continua) behave like transient interporosity flow (i.e., the KZ model).  They presented series of coefficients for a matrix memory kernel, which are mathematically equivalent to diffusion into matrix blocks of different geometry through analogies between the series solution and analytical solutions (e.g., similar to those for dual porosity by \citet{swaan76} and \citet{najurieta80}).  Diffusion into a slab is equivalent to an infinite distribution of pseudo-steady-state matrix domains given by
\begin{equation}
  \label{eq:haggerty-alpha}
  u_i = \frac{(2i-1)^2 \pi^2 \epsilon_i \kappa_i}{4 \omega_i }  \quad i=0,1, \dots
\end{equation}
and 
\begin{equation}
  \label{eq:haggerty-beta}
  \hat{\chi}_i = \frac{8 \omega_i}{(2i-1)^2 \omega_f \pi^2}  \quad i=0,1, \dots.
\end{equation}
\cite{haggerty1995} presented similar series of coefficients representing diffusion into cylinders or spheres (as did \citealp{swaan76}).  For their problem, they found truncating the series at $N=100$ resulted in pseudo-steady-state solutions within 1\% of matrix diffusion, with errors confined to early time.  The series are quick to compute, and using 1000 or more terms is not computationally expensive. 

Alternative distributions of matrix continua and properties can be specified by other widely used probability distributions, such as  exponential, normal, lognormal, or linear (e.g., \citealp{ranjbar2012one}; \citealp{malama13core}).

\section{Solutions}
The solution to (\ref{eq:multirate-nonD-fracture2}) in cylindrical coordinates can be found directly in Laplace space, analogous to solutions in the literature for traditional dual-porosity problems \citep{warren1963behavior,kazemi1969spe,mavor79} for either Type I (specified down-hole pressure) or Type II (specified flowrate) wellbore boundary conditions.  

\subsection{Flow problem of interest}
The following assumptions and boundary conditions are used to develop Laplace-space analytical solutions to the governing equations derived in the previous section:
\begin{enumerate}
\item single-phase slightly compressible flow (i.e., water, brine, oil, or gas treated using appropriate pseudo-variable methods),
\item a specified flowrate $Q(t)$ (Type II) or pressure change $\Delta p(t)$ (Type I) at the well completion,
\item the completion only intersects or interacts with fractures, with the matrix connected to the completion through fractures,
\item a symmetry no-flow boundary conditions parallel to the fracture midway between two fractures, and
\item a Type-III boundary condition at far edge of the domain, $r=R$ (Figure~\ref{fig:kazemi-slab}).  The Type-III boundary condition can either represent no-flow  (bounded reservoirs), specified  head (a circular island or  laboratory sand tank with a specified head condition), or a linear combination of the two.
\end{enumerate}

The pressure solution is developed for drawdown, $\psi = \left( p-p_0\right)/P_c$  (change from an initial state), therefore only the homogeneous initial condition is considered. 

The Laplace-transformed governing flow equation (\ref{eq:multirate-nonD-fracture2}) in cylindrical coordinates with radial symmetry is the modified Helmholtz equation for radial coordinates, 
\begin{equation}
  \label{eq:radial-multirate-nond}
\frac{\partial^2 \overline{\psi}_{f}}{\partial r_D^2} + \frac{1}{r_D} \frac{\partial \overline{\psi}_{f}}{\partial r_D} - \overline{\psi}_{f} \eta^2 = 0,
\end{equation}
where $r_D=r/L_c$ and $\eta^2 = s \omega_f \left(1 + \bar{g}\right)$ is a purely imaginary wave number, also called the fracture function \citep{al-ahmadi11}.  We pick $L_c=r_w$ as a characteristic length for radial flow to a well.

\subsection{Approach}
Analogous to \cite{mavor79}, we compute time-domain  numerical values from solutions to (\ref{eq:radial-multirate-nond}) using a numerical inverse Laplace transform algorithm \citep{de1982improved}. Several feasible alternative numerical inversion approaches exist \citep{kuhlman2013}, some which only require real values of $s$; the \cite{de1982improved} approach requires complex $s$.   

We present solutions for both finite and infinite domains as well as for specified bottomhole pressure or flowrate using modified Bessel functions --  the radial eigenfunctions of the modified Helmholtz equation in cylindrical coordinates.  

The multiporosity conceptualization may be appropriate for flow in fractured rock in which hydraulic fracturing is being performed, but the simple homogeneous cylindrical geometry of the radial solution presented here cannot adequately represent pumping discrete fractures under typical field conditions \citep{gringarten1982flow,clarkson13review}. The governing equations presented here could be solved for discrete linear fractures using analytical solutions for single fractures (e.g., \citealp{gringarten74}) or using analytic element superposition-based combinations of line element solutions \citep{bakker2011,biryukov12}. These approaches can better represent the geometry encountered during hydraulic fracturing.  Line sinks and sources, or narrow ellipses and polygons of high permeability can be used to represent discrete fractures added to a uniformly fractured multiporosity rock.

\subsection{Infinite Domain Solutions}
\label{sec:inf-domain-solution}
For an infinite radial domain we assume a homogeneous far-field Type-I boundary condition, $\lim_{r_D \rightarrow \infty} \overline{\psi}_{f}(r_D) = 0$, while the wellbore boundary condition is specified in two different ways as follows.

For specified down-hole flowrate, we choose $P_c = \mu B q/(2\pi h_f k_{f})$, where $Q(t) = q\bar{f}_t$ is a volume flowrate at the surface, $q$ is a constant characteristic flowrate, $h_f$ is the portion of the well completion open to fractures, and $B$ is a dimensionless formation volume factor correcting for the difference between down-hole and surface flowrates due to compressibility (in low-pressure or incompressible flow systems $B=1$).  This choice of $P_c$ results in the wellbore boundary condition    
\begin{equation*}
\left.  \frac{\partial \overline{\psi}_{f}}{\partial r_D} \right|_{r_D=1} = - \bar{f}_t ,
\end{equation*}
where $\bar{f}_t$ is the Laplace-transformed temporal behavior of the boundary condition (i.e., temporal fluctuations from $q$).  

Using $\bar{f}_t$, we consider general time behavior using the Laplace-space form of Duhamel's theorem \citep[Chap.~5]{ozisik1993}. For example, a step function on at $t_D=t_{D0}$ is $\bar{f}_t=e^{-t_{D0} s}/s$ (with the typical case of $\bar{f}_t=1/s$ for $t_{D0}=0$), and a pulse function during $t_{D0} \le t_D \le t_{D1}$ is $\bar{f}_f=\left(e^{-t_{D0} s} - e^{-t_{D1} s}\right)/s$. The temporal behavior of an arbitrary flowrate, $Q(t)$, or down-hole pressure, $\Delta p(t)$, is readily approximated using piecewise constant or linear behavior (e.g., \citealp{streltsova88} or \citealp{mishra13}).  For nearly  constant  or monotonically declining specified pressures and flowrates, the numerical inverse Laplace transform algorithm converges quickly (e.g., 21 Laplace-space function evaluations for each time is typical).   Many more Laplace-space function evaluations are needed to ensure accuracy when using periodic or rapidly fluctuating rates \citep{kuhlman2013}.  For this reason, we recommend fitting $\bar{f}_t$ to follow general observed trends, rather than every observed fluctuation. 

The Laplace-space solution to (\ref{eq:radial-multirate-nond}) for $\bar{\psi}_f$ in the completion $(r_D=1)$, given this boundary condition at the wellbore, is
\begin{equation}
  \label{eq:infinite-solution}
  \bar{\psi}_f^{(q,\infty)} = -\bar{f}_t \frac{\mathrm{K}_0(\eta)}{\eta \mathrm{K}_1(\eta)},
\end{equation}
where $\mathrm{K}_n(x)$ is the second-kind modified Bessel function of order $n$ \citep[\S10]{NIST:DLMF}.  

We do not consider wellbore storage explicitly.  \cite{mavor79} presented similar solutions including both wellbore storage and skin effects; these solutions could be used with the multiporosity conceptualization to approximately include wellbore storage and skin effects.  To rigorously include wellbore storage effects would additionally require including terms related to the time derivative of $\bar{f}_t$.

For specified downhole pressure, we chose $P_c(t_D)=p_0 \bar{f}_t$.  The Laplace-space solution to (\ref{eq:radial-multirate-nond}) for flowrate in the completion, given the Type-I boundary condition at the wellbore, is
\begin{equation}
  \label{eq:pressure-infinite-solution-flux}
  \bar{q}^{(P,\infty)} = \pi h_{fD}  \eta \bar{f}_t \frac{\mathrm{K}_1(\eta)}{\mathrm{K}_0(\eta)}.
\end{equation}

\subsection{Finite Domain Solutions}
For a finite cylindrical domain centered on the well completion, with a homogeneous Type-III  boundary condition at $r=R$ ($r_D=R_{D}$), the Laplace-space solution to (\ref{eq:radial-multirate-nond}) is given generally by \citet[\S13.4]{carslaw59}.  The specific Laplace-space solution to (\ref{eq:radial-multirate-nond}) for $\bar{\psi}_f$  in the completion, given the specified flowrate boundary condition at the wellbore and the Type-III  boundary condition $\frac{k_f}{\mu} \frac{\partial \Delta p_f}{\partial r} + H \Delta p_f = 0$ at  $r=R$ (nondimensionalized in Laplace space to $\frac{\partial \bar{\psi}_f}{\partial r_D} + H_D \bar{\psi}_f = 0$, where $H_D=H \mu L_c/k_f$), is
\begin{equation}
  \label{eq:finite-solution}
  \bar{\psi}_f^{(q,R)} = \frac{\bar{f}_t}{\eta} \frac{\mathrm{I}_0(\eta) \xi  + \mathrm{K}_0(\eta) \zeta }{\mathrm{I}_1(\eta)\xi  - \mathrm{K}_1(\eta)\zeta}
\end{equation}
where 
\begin{align*}
\xi=\eta \mathrm{K}_1(\eta R_{D}) &- H_D \mathrm{K}_0(\eta R_{D}),\\
\zeta= \eta \mathrm{I}_1(\eta R_{D}) &+ H_D \mathrm{I}_0(\eta R_{D}),
\end{align*} 
$H$ is the Type-III boundary surface conductivity, and $\mathrm{I}_n(x)$ is the first-kind modified Bessel function of order $n$ \citep[\S10]{NIST:DLMF}. 

The Laplace-space solution to (\ref{eq:radial-multirate-nond}) for flowrate,  given the specified pressure boundary condition at the wellbore and a homogeneous Type-III boundary condition  at the outer boundary, is
\begin{equation}
  \label{eq:pressure-finite-solution-flux}
  \bar{q}^{(P,R)} = \pi h_{fD} \eta \bar{f}_t  \frac{I_1(\eta)\xi - K_1(\eta)\zeta}{I_0(\eta)\xi + K_0(\eta)\zeta}.
\end{equation}
In the limiting case where $H_D=0$, the boundary condition at $r_D=R_{D}$ becomes a no-flow condition.  When $H_D \rightarrow \infty$, the outer boundary condition becomes a specified-pressure condition.  In the more general case, $H_D$ represents ``leakiness'' of the outer boundary; a second reservoir beyond $r_D=R_D$ is providing a non-zero flux into the domain, proportional to the change in pressure at the boundary.

\section{Model Behavior}
We present examples of some typical behavior of the multiporosity model. First we illustrate how the series of matrix porosities (Equations \ref{eq:haggerty-alpha} and \ref{eq:haggerty-beta}) can approximate both WR and KZ flow as end members for the infinite domain solution.  Second, we illustrate model predictions for a range of different matrix and fracture properties. Finally, we illustrate the nature of the different boundary conditions ($H \in \{ 0,H,\infty \}$) at the radial extent of the domain  $r_D=R_{D}$. 

\begin{figure} % Figure 2
  \noindent\includegraphics[width=\textwidth]{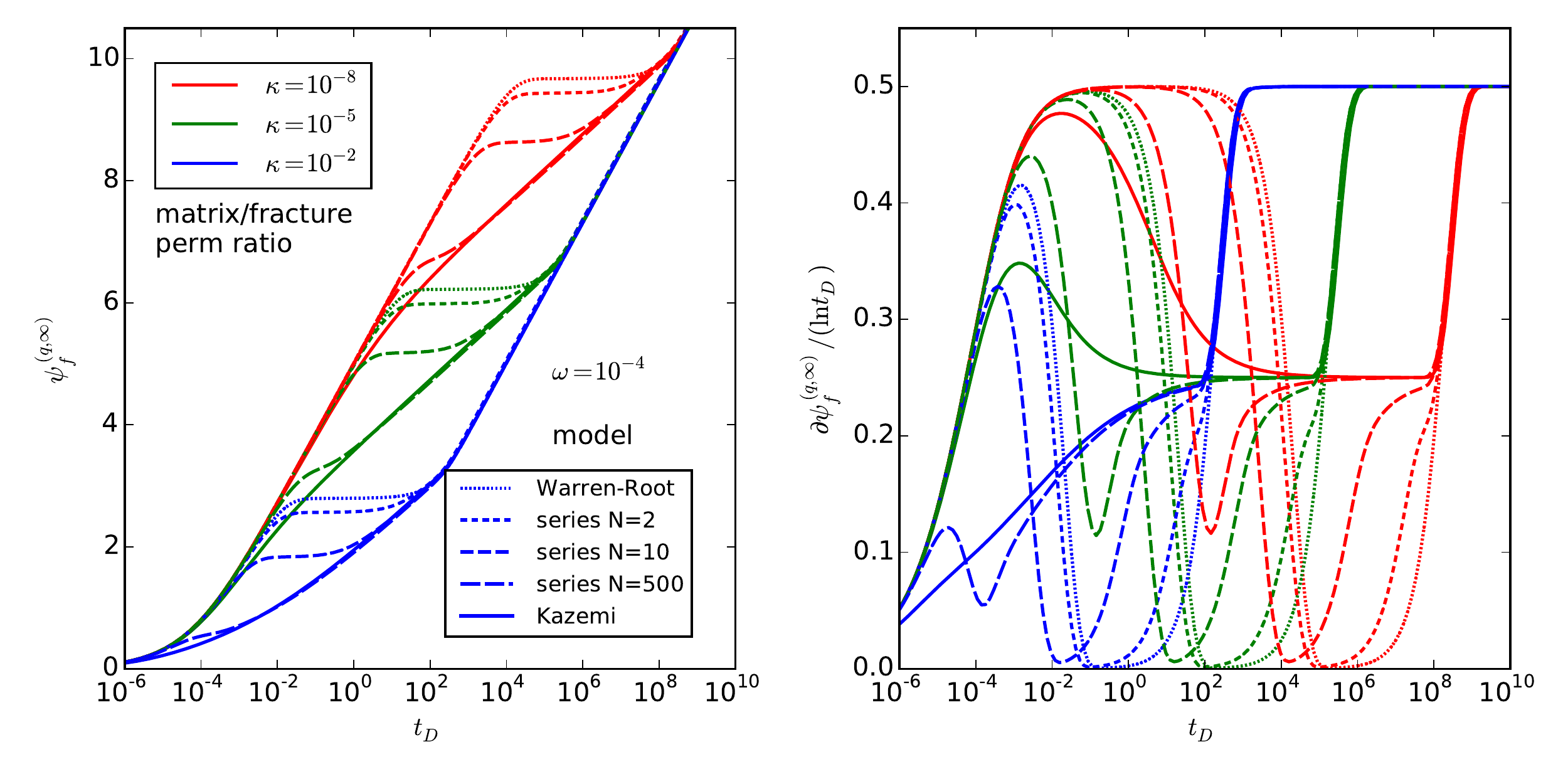}
  \caption{Multiporosity flow solution for constant downhole flowrate ($\bar{f}_t=1/s$) infinite domain, showing
approximation to WR ($N=1$), and KZ ($N \rightarrow \infty$) solutions.}
  \label{fig:WR-KZ-multiporosity-compare-specQ}
\end{figure}

For constant matrix properties, the multiporosity model can reproduce both pseudo-steady-state (WR) and transient dual-porosity (KZ) flow, as well as  a range of intermediate behaviors.  Figure \ref{fig:WR-KZ-multiporosity-compare-specQ} shows the behavior of the infinite domain solution (\ref{eq:infinite-solution}) for $u_j=u$, $\beta_j=\beta$, and different values of $N$. Using constant formation properties for an infinite number of pseudo-steady-state matrix porosities results in an equivalent transient KZ solution with similar properties. The figure shows the increase in dimensionless pressure drawdown due to production at a constant rate. The line types indicate the model, while the  line colors indicate the matrix/fracture permeability ratio $\kappa$. The right plot in Figure~\ref{fig:WR-KZ-multiporosity-compare-specQ} shows the slope with respect to $\ln(t)$, illustrating how the KZ model (solid line) has an intermediate slope which is 1/2 of its slope at early or late time.  The WR model has essentially  zero slope at intermediate time.  As more terms are added to the multiporosity model, the slope increases from zero to 1/2 the late time slope, beginning from the later-time portion of the intermediate time portion and moving back towards earlier time.

\begin{figure} % Figure 3
  \noindent\includegraphics[width=\textwidth]{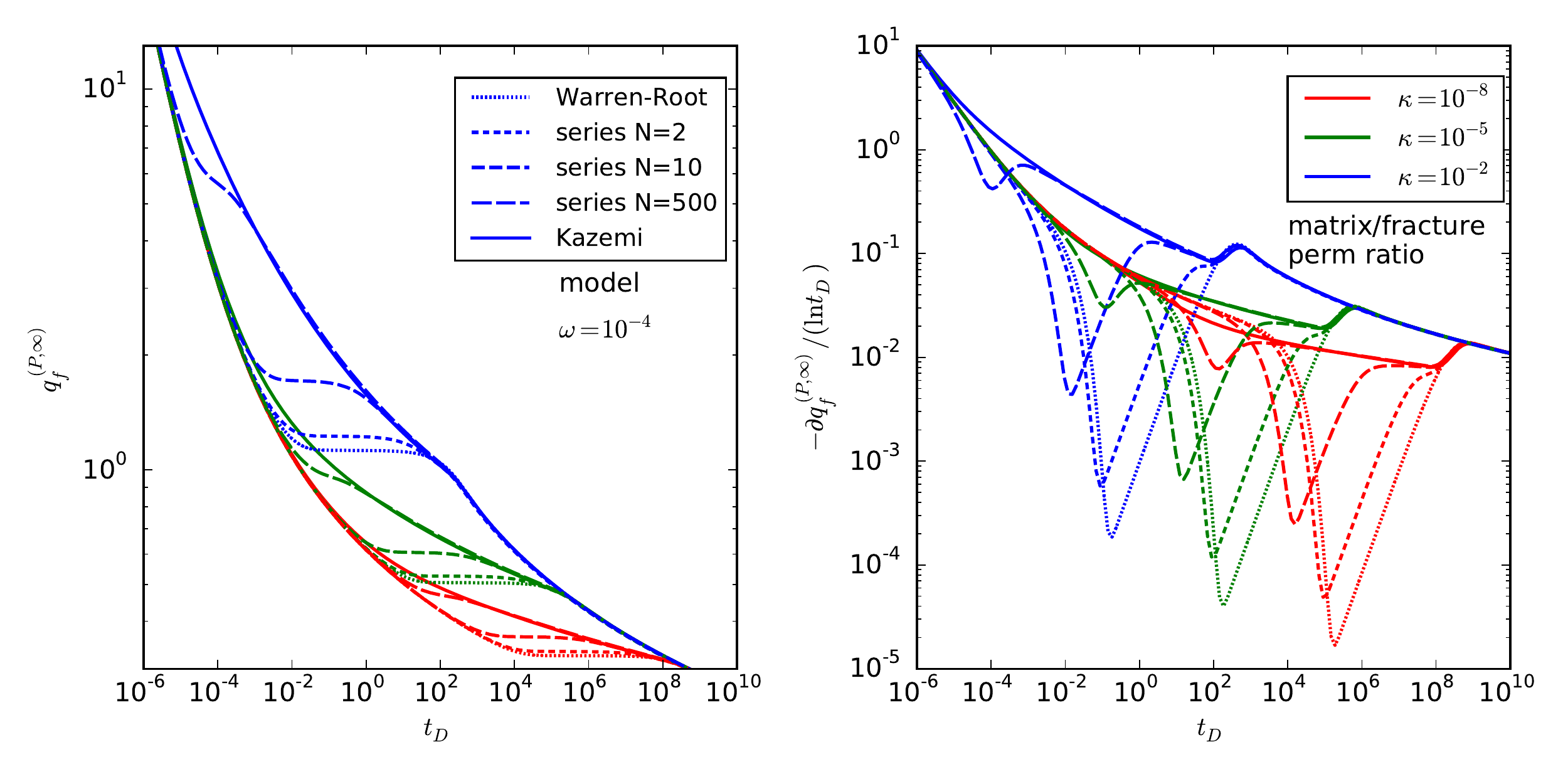}
  \caption{Multiporosity flow solution for constant downhole  pressure ($\bar{f}_t=1/s$) infinite domain, showing
approximation to WR ($N=1$), and KZ ($N \rightarrow \infty$) solutions.}
  \label{fig:WR-KZ-multiporosity-compare-specP}
\end{figure}

Figure~\ref{fig:WR-KZ-multiporosity-compare-specP}  shows the infinite domain  solution for flowrate, given a specified bottomhole pressure (\ref{eq:pressure-infinite-solution-flux}).  This figure shows the decay in flowrate, due to production at a specified bottomhole pressure, illustrating the variation between the WR and KZ endmembers.

\begin{figure} % Figure 4
  \noindent\includegraphics[width=0.5\textwidth]{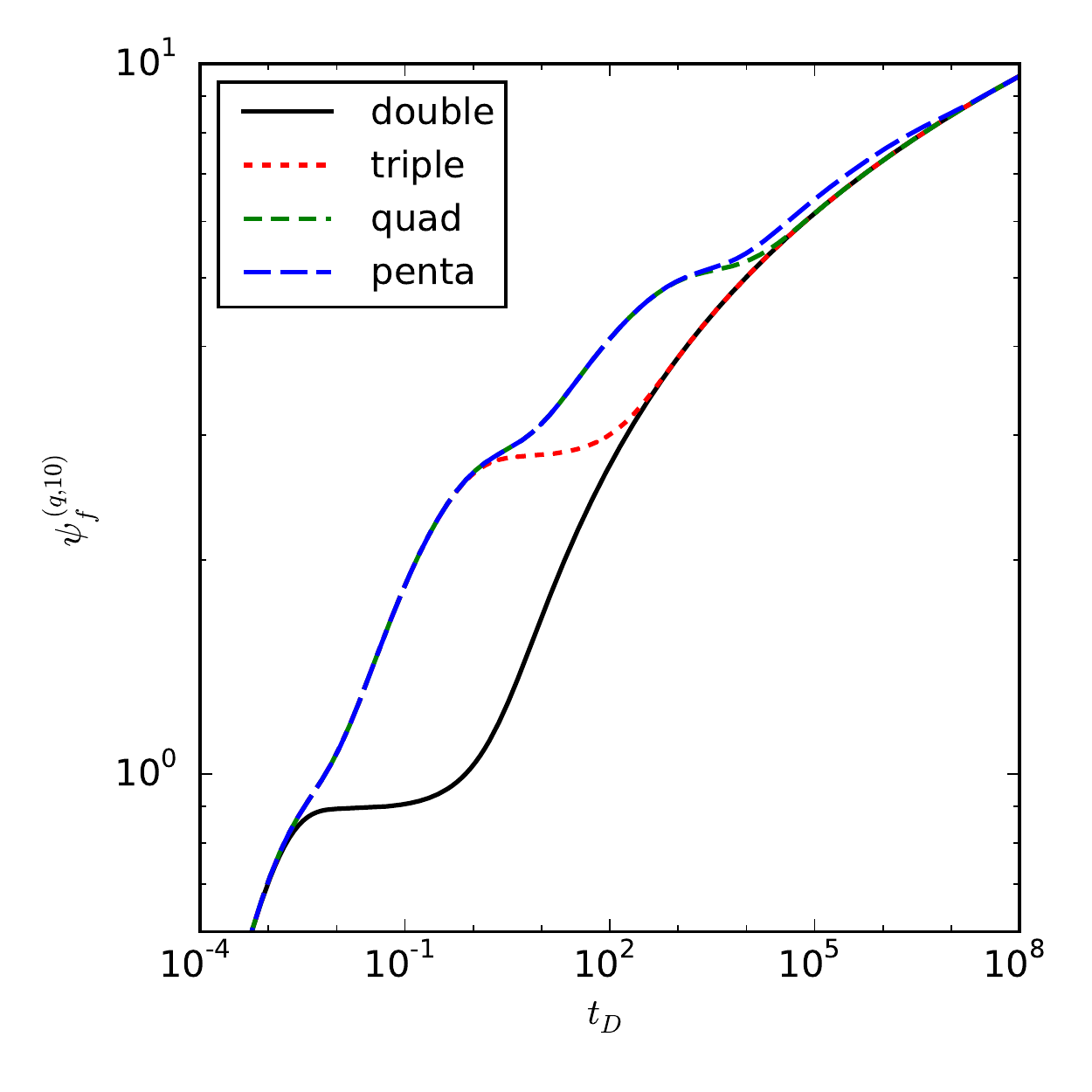}
  \caption{Multiporosity flow solution for constant flowrate ($\bar{f}_t=1/s$) finite domain (Type II, $H_D=0$), multiple porosity solutions with different matrix properties ($\kappa_j$ and $\omega_j$) in each matrix continuum.}
  \label{fig:double-triple-quad-penta}
\end{figure}

Figure~\ref{fig:double-triple-quad-penta} shows multiporosity solutions with different matrix properties for each matrix continuum, rather than uniform properties, as shown in Figures \ref{fig:WR-KZ-multiporosity-compare-specQ} and \ref{fig:WR-KZ-multiporosity-compare-specP}.  These solutions are analogous to classical double- and triple-porosity models, where the two matrix permeabilities represent ``good'' or ``bad'' quality rock \citep{clossman75}, or possibly heterogeneity in the matrix system \citep{al-ahmadi11}. Table~\ref{tab:parameters} shows the values of $u_j$ and $\hat{\chi}_j$ used in (\ref{eq:g}). These values are derived from $\omega_f=0.001$ and $\kappa_j = \{0.5, 0.05, 5\times10^{-5}, 5\times10^{-8}\}$.   Similar to Figure~\ref{fig:WR-KZ-multiporosity-compare-specQ} for constant parameters across the different porosities, the flat transition regions beyond the first transition disappear with increasing number of matrix continua for this selection of matrix properties.  The double-porosity model has one transition zone, the triple-porosity model has two, and the quad-porosity has three transition periods (a single porosity model has no transitions).

\begin{figure} % Figure 5
  \noindent\includegraphics[width=\textwidth]{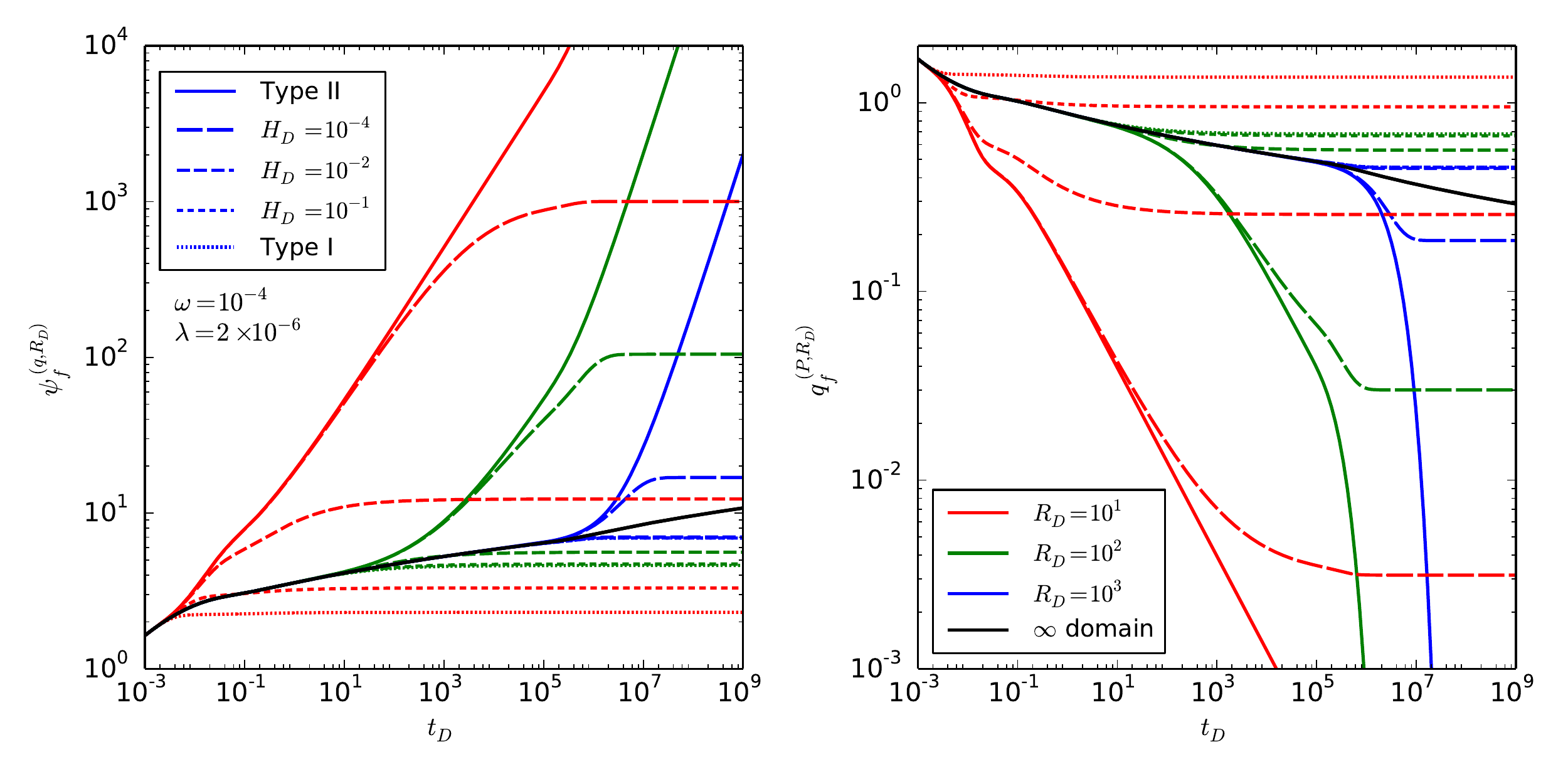}
  \caption{Multiporosity flow solution (KZ, $N \rightarrow \infty$) for specified flowrate (left) and downhole pressure (right) for a range of domain sizes and $H_D$ (Type I: $H_D=10^{8}$; Type II: $H_D=0$).}
  \label{fig:outer-boundary-conditions}
\end{figure}

Figure~\ref{fig:outer-boundary-conditions} shows the effects of boundary conditions at the radial extent of the domain. The solid curves represent the solution for $H=0$ (a no-flow boundary condition).  For the smaller domain  (red lines), the effects of the boundary are seen before the effects of the matrix porosity.  In larger domains (green and blue), the effects of the boundary aren't observed until later time.  The black curve shows the solution for the infinite domain.  The dotted lines represent the solution for $H  \rightarrow \infty$ (implemented as $10^{8}$), which is effectively a no-drawdown condition at the edge of the domain.  

Type I and III boundary conditions are less applicable to bounded geologic reservoirs, but these solutions may be useful in other applications, where a high-permeability reservoir surrounds a double- or multiple-porosity medium.  The linear relationship between flux and change in pressure between the domain and boundary condition at $R_{D}$ is analogous to the pseudo-steady-state assumption for flux between the fracture and matrix.

\section{Summary}
The multiporosity model is shown to be a generalization of pseudo-steady-state double and triple porosity solutions of \cite{warren1963behavior}, \cite{clossman75}, and \cite{ci1983exact}, as well as the transient double porosity solution of \cite{kazemi1969spe}, given different distributions of matrix properties.  Previous solutions for interporosity flow with variable block sizes \citep{ranjbar2012one} have not included the connection to the solution of \cite{kazemi1969spe}, and have focused on the block-size distribution properties.

The cylindrically symmetric solutions presented here for continuously fractured and discretely fractured systems show how the multiporosity model can be presented as a unifying approach to solve multiple flow regimes using a single diffusion-based model, when the same properties are used across the different matrix continua.  When different properties are used in each matrix porosity, the solution can produce results based upon the interaction of multiple physical reservoirs (i.e.,  macro- and micro-fractures). The radially symmetry Laplace-domain solutions presented are given for specified downhole pressure and flowrate, with finite or infinite domains, and a general Type-III boundary condition at the far extent of the domain.

Although there have already been many approximate and analytical solutions proposed for flow in fractured rock \citep{gringarten1982flow,chen1989transient}, we propose the multiporosity solution as a simple yet unifying approach to include the effects of multiple interacting reservoirs. The conceptual solute transport approach of \cite{haggerty1994,haggerty1995,haggerty1998} has been adapted to the diffusion of pressure between fracture and matrix reservoirs.

\section*{acknowledgments}
This research was funded by Sandia Laboratory-Directed Research and Development.  Sandia National Laboratories is a multi-program laboratory managed and operated by Sandia Corporation, a wholly owned subsidiary of Lockheed Martin Corporation, for the U.S. Department of Energy's National Nuclear Security Administration under contract DE-AC04-94AL85000.

A Fortran program implementing the multiporosity model is available from the corresponding author.
%%\end{acknowledgments}

%\bibliographystyle{abbrvnat}
%\bibliography{dual-porosity}

%%\newpage

% ----------------------------------------

\begin{table}  % Table 1
  \caption{Parameters used in multiporosity model shown in Figure~\ref{fig:double-triple-quad-penta}}\label{tab:parameters}
  \centering
  \begin{tabular}{c c l l l}
    \hline
    Model & N & $\omega_j$ & $u_j$ & $\hat{\chi}_j$ \\
    \hline
    Double & 1 & \{0.999\}& \{0.5\} & \{999\} \\
    Triple & 2& \{0.005, 0.994\}&\{100, 0.005\} & \{5, 994\} \\
    Quad & 3& \{0.005, 0.05, 0.944\}&\{100, 0.1, 5$\times10^{-5}$\} & \{5, 50, 944\} \\
    Penta & 4& \{0.005, 0.05, 0.5, 0.444\} &\{100, 0.1, $10^{-4}$, $10^{-7}$\} & \{5, 50, 500, 444\} \\
    \hline
  \end{tabular}
\end{table}

\begin{table}  % Table 2
  \caption{Fundamental quantities}\label{tab:nomenclature}
  \centering
  \begin{tabular}{r l c}
    \hline
    $B$ & formation volume factor (formation volume/surface volume) & - \\
    $c$ & total system compressibility &  L$^{-1}$ \\
    $c_{\{ f,j\}}$ & fracture and $j$th matrix domain compressibilities &  L$^{-1}$ \\
    $h_f$ & length of well completion open to fractures &  L \\
    $H$ & Type-III boundary condition surface conductivity at $R$ & T$^{-1}$ \\
    $p_{\{f,j\}}$ & fracture and $j$th matrix domain pressure heads &  L \\
    $p_0$ & initial pressure head &  L \\
    $k_{\{f,j\}}$ & fracture and $j$th matrix domain permeabilities &  L$^2$ \\
    $L$ & half width of matrix domain blocks &  L \\
    $L_c$ & characteristic length &  L \\
    $P_c$ & characteristic pressure head &  L \\
    $Q$ & time-variable volumetric flowrate at surface &  L$^3 \cdot$ T$^{-1}$ \\
    $q$ & constant volumetric flowrate at surface &  L$^3 \cdot$ T$^{-1}$ \\
    $r_w$ & wellbore radius &  L \\
    $R$ & domain radius (when finite) &  L \\
    $s$ & Laplace transform parameter & - \\
    $t$ & time since beginning of testing or production & T \\
    $T_c$ & characteristic time & T \\
    $\phi$ & total system porosity & - \\
    $\phi_{\{f,j\}}$ & fracture and $j$th matrix domain porosities & - \\
    $\mu$ & fluid viscosity &  L$\cdot$T \\
    $\chi$ & interporosity flow parameter probability mass function & - \\
    \hline
  \end{tabular}
\end{table}

\begin{table}  % Table 3
  \caption{Derived dimensionless quantities}\label{tab:dimensionless}
  \centering
  \begin{tabular}{rcll}
    \hline
    $h_{fD}$ &=& $h_f/L_c$ & dimensionless fracture completion length \\
    $H_D$ &=& $H \mu L_c/k_f $ & dimensionless boundary conductivity \\
    $L_{D}$ &=& $L/L_c$ & dimensionless matrix domain block length \\
    $r_D$ &=& $r/L_c$ & dimensionless fracture coordinate (perpendicular to wellbore) \\
    $t_D$ &=& $t/T_c$ & dimensionless time \\
    $R_{D}$&=& $R/L_c$ & dimensionless domain radius\\
    $y_D$ &=& $y/L_c$ & dimensionless coordinate parallel to wellbore \\
    $\beta_j$ &=& $\omega_j/\omega$ & storage capacity ratio for matrix domain $j$ \\
    $\eta$ &=& $\sqrt{s \omega_f \left( 1 + \bar{g}\right) }$ & Helmholtz wave number for multiporosity flow \\
    $\kappa_j$ &=& $k_j/k_f$ & matrix fracture permeability ratio for matrix domain $j$ \\
    $\omega_{\{f,j\}}$ &=& $\phi_{\{f,j\}} c_{\{f,j\}} / \left( \phi c\right)$ & fracture and $j$th matrix domain storage ratio \\
    \hline
  \end{tabular}
\end{table}

\end{document}